\documentclass[conference]{IEEEtran}
\IEEEoverridecommandlockouts
\usepackage{cite}
\usepackage{amsmath,amssymb,amsfonts}
\usepackage{algorithmic}
\usepackage{graphicx}
\usepackage{textcomp}
\usepackage{xcolor}
\def\BibTeX{{\rm B\kern-.05em{\sc i\kern-.025em b}\kern-.08em
    T\kern-.1667em\lower.7ex\hbox{E}\kern-.125emX}}
    
\usepackage{tikz}
\usepackage{pgfplots}
\usetikzlibrary{spy}

\def\*#1{\mathbf{#1}}
\newcommand{\vect}[1]{{\mathbf{#1}}}
\newcommand{\mat}[1]{{\mathbf{#1}}}

\newcommand{\TS}[1]{\mathcal{T}_{#1}}
\newcommand{\RS}[1]{\mathcal{T}_{#1}}
\renewcommand{\sf}[1]{\mathsf{#1}}
\renewcommand{\intercal}{{T}}
\newcommand{\PMDoF}{\textnormal{PMDoF}}

\newcommand{\lo}[1]{{#1}}
\newcommand{\sh}[1]{{\color{green}#1}}
\renewcommand{\sh}[1]{}
\newif\ifADDpagenumber
\ADDpagenumbertrue

\allowdisplaybreaks[4]
\usepackage{amsthm}
\newtheorem{theorem}{Theorem}
\newtheorem{lemma}{Lemma}
\newtheorem{corollary}{Corollary}
\newtheorem{remark}{Remark}

\begin{document}

\title{DoF of a Cooperative  X-Channel with an Application to Distributed Computing}


\author{
    \IEEEauthorblockN{Yue Bi\IEEEauthorrefmark{1}\IEEEauthorrefmark{2}, Mich\`ele Wigger\IEEEauthorrefmark{1}, Philippe Ciblat\IEEEauthorrefmark{1}, Yue Wu\IEEEauthorrefmark{2}}
    \IEEEauthorblockA{\IEEEauthorrefmark{1} \textit{LTCI, Telecom Paris, IP Paris}, 91120 Palaiseau, France\\
    \{bi, michele.wigger, philippe.ciblat\}@telecom-paris.fr}
    \IEEEauthorblockA{\IEEEauthorrefmark{2}\textit{School of Electronic Information and Electrical Engineering, Shanghai Jiao Tong University}, China\\ 
    wuyue@sjtu.edu.cn}
}

\maketitle

\ifADDpagenumber
\thispagestyle{plain} 
\pagestyle{plain}
\fi

\begin{abstract}
 We consider a cooperative X-channel with $\sf K$ transmitters (TXs) and $\sf K$ receivers (Rxs) where Txs and Rxs are gathered into groups of size $\sf r$ respectively. Txs belonging to the same group cooperate to jointly transmit   a  message to each of the $\sf K- \sf r$ Rxs in all other groups, and each Rx individually decodes all its intended messages. By introducing a new interference alignment (IA) scheme, we prove that when  $\sf K/\sf r$ is an  integer the sum Degrees of Freedom (SDoF) of this channel is lower bounded by $2\sf r$ if $\sf K/\sf r \in \{2,3\}$ and by $\frac{\sf K(\sf K-\sf r)-\sf r^2}{2\sf K-3\sf r}$ if $\sf K/\sf r  \geq 4$. We also prove that the SDoF is upper bounded by  $\frac{\sf K(\sf K-\sf r)}{2\sf K-3\sf r}$. The proposed IA scheme finds application  in a wireless distributed MapReduce framework, where  it improves the normalized data delivery time (NDT)  compared to the state of the art.
\end{abstract}


\section{Introduction}
Identifying the capacity region of a multi-user channel with interference is generally a difficult task. One way to provide insights on the capacity region is  to resort to the {\it sum degrees of freedom (SDoF)} of the channel, which characterizes the pre-log approximation of the sum-capacity in the asymptotic regime of infinite Signal-to-Noise Ratios (SNR)  \cite{Algoet}, i.e., when the network operates in the interference-limited regime. The  study of  the SDoF of interference channels (IC) and X-channels (where each Tx sends a message to each Rx) with and without cooperation has a rich history, see e.g., \cite{foschini1998limits, lapidoth_cognitive_2007, devroye2007multiplexing,  cadambe_interference_2008,jafar_degrees_2008,cadambe_interference_2009,annapureddy_degrees_2012,cadambe_can_2008, motahari_real_2014, zamanighomi_degrees_2016, wei_iterative_2019}. In particular,  it has been shown that  the SDoF of a  fully-connected  $\sf K$-user IC without cooperation  is $\sf K/2$ when the channel coefficients are independent and identically distributed (i.i.d.) fading according to a continuous distribution \cite{jafar_degrees_2008} and the SDoF of the corresponding X-channel is  ${\sf K}^2/(2{\sf K}-1)$. Both these SDoFs are  achieved with interference alignment (IA) \cite{cadambe_can_2008}.

\begin{figure}[ht]
    \centering
    \includegraphics[width=0.4\textwidth]{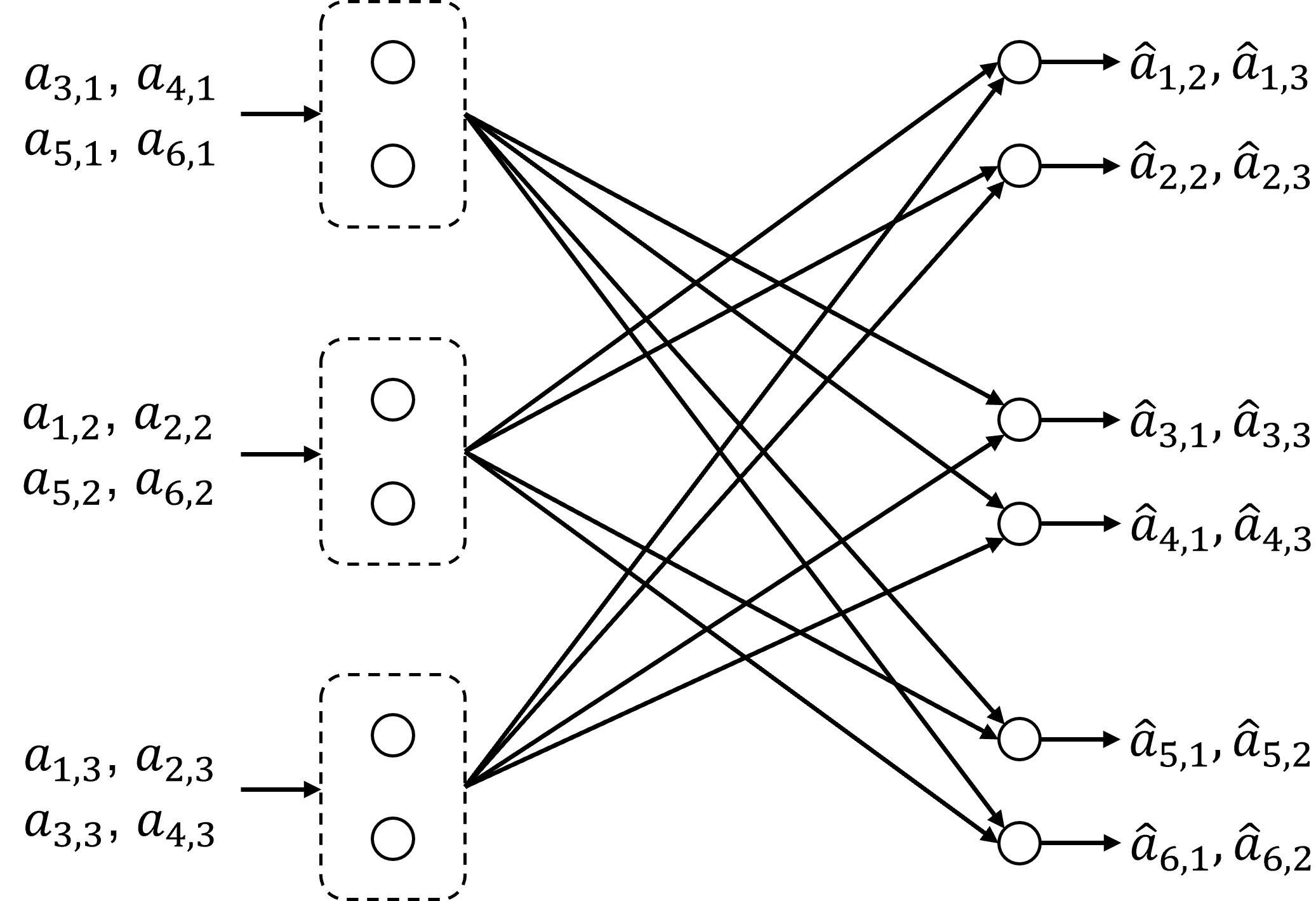}
    \caption{The cooperative X-channel model for $\sf K=6$, $\sf r=2$}
    \label{fig:channel_model}
    \vspace{-3mm}
\end{figure}

In this paper, we study   the SDoF of the partially-connected  X-channel in Fig.~\ref{fig:channel_model}. Txs/Rxs are gathered into groups of $\sf r>0$ consecutive  Txs/Rxs, and each Rx observes a linear combination of all Tx-signals in Gaussian noise, except for the signals sent by its corresponding Tx-group. Txs in the same group cooperate to jointly transmit a message to each Rx in all other groups, while Rxs decode their intended messages independently of each other.
For this network model, we propose a new IA scheme that achieves SDoF  $(\sf K(\sf K-\sf r)-\sf r^2)/(2\sf K-3\sf r)$ whenever the ratio of $\sf K$ by $\sf r$ is an integer larger than $3$. This lower bound improves over the $\sf{K}/2$ lower bound in  \cite[Theorem 2]{cadambe_can_2008} which was reported for the special case ${\sf r}=1$.  We further show a SDoF upper bound of 
$\sf K (\sf K - \sf r)/(2\sf K - 3 \sf r)$.

The motivation for studying our particular X-channel stems from an application in \emph{MapReduce distributed computing (DC)}. MapReduce  is a popular framework to carry out heavy computation tasks and runs in three phases \cite{ng_comprehensive_2021, dean2008mapreduce}. In the first \emph{map phase}, nodes calculate intermediate values (IVA) from  their associated input files. In the subsequent \emph{shuffle phase}, nodes exchange these IVAs to obtain all IVAs  required to run the final \emph{reduce phase} where they compute the desired output function. The largest part of the execution time in  MapReduce systems stems from the IVA \emph{delivery time} during the shuffle phase. Several works proposed to reduce this delivery time through smart coding. More specifically, in wired networks, delivery time is decreased by sending appropriate linear combinations of the IVAs \cite{li_fundamental_2018, yan2019storage, xu_new_2021}, over wireless cellular networks \cite{li_scalable_2017, yang_data_2019} a similar effect is achieved through simple interference cancellation at the receiving nodes, and over wireless interference networks a gain was achieved by zero-forcing \cite{li_wireless_2019}. In this paper, we show further improvement in wireless interference networks using a novel IA scheme.


\textit{Notations:} We use sans serif font for constants, bold for vectors and matrices, and calligraphic font for most sets. The sets of complex numbers and positive integers are denoted $\mathbb{C}$ and $\mathbb{Z}^+$.  For a finite set $\mathcal{A}$, let $|\mathcal{A}|$ denote its cardinality. For any $n\in\mathbb{Z}^+$,  define $[n] \triangleq \{1,2,\ldots,  n\}$ and let $\mat{Id}_n$ be  the $n \times n$ identity matrix. Let further $\boldsymbol{0}$  denote the  all-zero matrix and    $\boldsymbol{1}$ the all-one vector, where the dimensions will be clear from the context.  For any vector $\vect{v}$, let $\textnormal{diag}(\vect{v})$ be the diagonal matrix with diagonal entries given by the elements of the vector $\vect{v}$.  
When writing $[ \boldsymbol{v}_i\colon i \in \mathcal{S}]$ or $[\boldsymbol{v}_i]_{i\in \mathcal{S}}$ we mean the matrix consisting of the set of columns $\{\vect{v}_i\}_{i\in\mathcal{S}}$. For two matrices $\*A$ and $\* B$, we use $\* A \otimes \* B$ to denote  their Kronecker product. 

\section{Channel model}\label{sec:channel_model}

Consider an interference network with $\sf  K$ Txs and $\sf K$ Rxs labeled from $1$ to $\sf K$. For a  given  group-size $\sf r \geq 1$, where  $\sf K$ is assumed divisible by $\sf r$, we define the group of Txs/Rxs
\begin{equation}\label{eq:Group}
\TS{k} \triangleq \{(k-1)\sf r +1,\ldots,  k \sf r \}, \qquad k \in[ \tilde{\sf K}],
\end{equation}
where $\tilde{\sf K}\triangleq \sf K/ \sf r$. 

In our network model, each Rx $p$ in Rx-group $\RS{j}$ observes a linear combination of the signals sent by all Txs \emph{outside}  Tx-group $\TS{j}$, corrupted by Gaussian noise.  Denoting Tx~$q$'s slot-$t$ input by $X_{q}(t) \in \mathbb{C}$ and Rx~$p$'s slot-$t$ output by $Y_{p}(t) \in \mathbb{C}$, the input-output relation of the network is:
\begin{IEEEeqnarray}{rCl}\label{eq:gen}
   Y_p(t) = \sum_{q  \in [\sf K]\backslash \TS{j} } H_{p,q}(t) X_{q}(t) + Z_{p}(t) , \qquad p \in \RS{j},
\end{IEEEeqnarray}
where the sequences of complex-valued channel coefficients $\{H_{p,q}(t)\}$  and standard circularly symmetric Gaussian noises $\{Z_{p}(t)\}$ are both i.i.d. and  independent of each other and of all other channel coefficients and noises. 
The real and imaginary parts of a coefficient $H_{p,q}(t)$ are  i.i.d. according to a given continuous distribution on some bounded interval $[-\sf H_{\max},\sf H_{\max}]$ and  are known by all terminals even before communication starts.

We consider a scenario where all Txs in Tx-group $\mathcal{T}_k$ cooperatively transmit an individual message $a_{p,k}$ to each Rx~$p\in[\sf K]\backslash \RS{k}$ \emph{outside} Rx-group $k$. When communication is of blocklength $\sf T$, this message is  uniformly distributed over $\left[2^{\sf T\sf R_{p,k}}\right]$, where $\sf R_{p,k}\geq 0$ denotes the rate of transmission, and it is independent of all other messages and of  all channel coefficients and noise sequences. 
As a consequence, Tx~$q\in[\sf K]$ produces its block of channel inputs $X^{(\sf T)}_q \triangleq (X_{q}(1), \ldots, X_q(\sf T))$ as 
\begin{equation}
X^{(\sf T)}_q = f_q^{(\sf T)}\left(\left\{a_{p,k}\colon k =\left \lceil \frac{q}{ \sf r}\right\rceil ,\; p \in [\sf K]\backslash \RS{k}\right\}\right)
\end{equation}
by means of an encoding function $f_q^{(\sf T)}$ on appropriate domains and so that the  inputs satisfy the  block-power constraint
 \begin{IEEEeqnarray}{rCl}\label{eq:power}
\frac{1}{\sf T} \sum_{t=1}^{\sf T}     \mathbb{E}\left[|X_{q}(t)|^2\right] \leq\sf  P, \qquad q\in[\sf K].
 \end{IEEEeqnarray}

 Given a power $\sf P>0$, the \emph{capacity region} $\mathcal{C}(\sf P)$ is defined as the set of all  rate tuples $(\sf R_{p,k} \colon k \in [\tilde{\sf K}], \; p \in [\sf K]\backslash \RS{k})$ so that  for each blocklength $\sf T$  there exist
 encoding functions $\{f_{q}^{(\sf T)}\}_{q \in [\sf K]}$ as described above and decoding functions $\{g_{p,k}^{(\sf T)}\}$ on appropriate domains producing the estimates 
\begin{equation}
\hat{a}_{p,k} = g_{p,k}^{(\sf T)}( Y_p(1),\ldots, Y_p(\sf T)), \quad k\in[\tilde{\sf K}], \; p\in[\sf K]\backslash \RS{k},
\end{equation}
in a way that the sequence of error probabilities 
\begin{equation} \label{eq:error_prob}
p^{(\sf T)}(\textnormal{error}) \triangleq  \textnormal{Pr}\bigg[ \bigcup_{ \substack{k\in[\tilde{\sf K}]}} \bigcup_{  p\in[\sf K]\backslash \RS{k} } \hat{a}_{p,k}\neq a_{p,k} \bigg]
\end{equation}
tends to 0 as the blocklength $\sf T \to \infty$.

Our main interest is in the \emph{Sum Degrees of Freedom (SDoF)}:
\begin{IEEEeqnarray}{rCl}
    \text{SDoF} \triangleq \varlimsup_{\sf P \rightarrow \infty} \sup_{ \vect{R} \in \mathcal{C}(\sf P)} \sum_{k \in [\tilde{\sf K}]}\; \sum_{p \in [\sf K]\backslash \RS{k}} \frac{  \sf R_{p,k}}{\log \sf P}.
\end{IEEEeqnarray}

\section{Main Results}\label{sec:main}
The main results of this paper  are new upper and lower bounds on the SDoF of the network described in the previous Section~\ref{sec:channel_model}. We restrict attention to $\sf K / \sf r >1$, because for $\sf r ={\sf K}$ the Rxs only observe noise and trivially $\textnormal{SDoF}=0$.

\begin{theorem}\label{thm:dof_bounds}
When $\sf K / \sf r$ is an integer strictly larger than $1$, the SDoF of the network in Section~\ref{sec:channel_model} is lower bounded as: 
\begin{IEEEeqnarray}{rCl} \label{eqn:dof_lower_bound}
    \textnormal{SDoF} \geq \textnormal{SDoF}_{\textnormal{Lb}}\triangleq  \left\{\begin{array}{ll} 2\sf r & \textrm{if }  \sf K/\sf r  \in \{2,3\}, \\  \frac{\sf K(\sf K-\sf r)-\sf r^2}{2\sf K-3\sf r} & \textrm{if }   \sf K/\sf r  \geq 4, \end{array} \right.
\end{IEEEeqnarray} 
and upper bounded as:
\begin{IEEEeqnarray}{rCl} \label{eqn:dof_upper_bound}
\textnormal{SDoF}   \leq \frac{\sf K(\sf K-\sf r)}{2\sf K-3\sf r}.
\end{IEEEeqnarray}
\end{theorem}

\begin{IEEEproof}
See Section \ref{sec:ach} for the  proof of the lower bound and \lo{Appendix~\ref{sec:converse}}\sh{\cite[Appendix~A]{arxiv2022}} for the  proof of the upper bound.
\end{IEEEproof}

\lo{For $\sf K/\sf r \geq 4$, the additive gap between the lower and upper bounds in \eqref{eqn:dof_lower_bound} and \eqref{eqn:dof_upper_bound}  is  $\frac{\sf r^2}{2\sf K-3\sf r}$.  This gap is decreasing in $\sf K$ and increasing in $\sf r$.}

For $\sf K/\sf r\in\{2,3\}$ the bounds \eqref{eqn:dof_lower_bound} and \eqref{eqn:dof_upper_bound}  match and yield: 
\begin{corollary}
For $\sf K/\sf r \in \{2,3\}$, we have 
$\textnormal{SDoF} = 2\sf r.$
\end{corollary}

 For $\sf r=1$, our lower bound  \eqref{eqn:dof_lower_bound} improves over the lower bound  SDoF $\geq \sf K/2$ reported in \cite{cadambe_can_2008} for all values of $\sf K$.

\begin{remark}\label{rem:PUDoF}
By the symmetry of the setup and standard time-sharing arguments, the bound in \eqref{eqn:dof_lower_bound} implies the following bound on the \emph{Per-Message DoF (\PMDoF)} 
\begin{IEEEeqnarray}{rCl}\label{eq:PUDoF}
    \PMDoF & \triangleq&  \lim_{\sf P \rightarrow \infty} \sup_{ \vect{R} \in \mathcal{C}(\sf P)} \min_{\substack{k \in [\tilde{\sf K}]\\ p \in [\sf K]\backslash \RS{k}}}\frac{ \sf R_{p,k} }{ \log \sf P}  \geq \frac{  \textnormal{SDoF}_{\textnormal{Lb}}}{\sf K(\sf K/\sf r-1)}. \IEEEeqnarraynumspace 
\end{IEEEeqnarray} 
\end{remark}

\lo{In the following Section \ref{sec:application_WDC} we discuss an application of our lower bound in \eqref{eqn:dof_lower_bound}  to DC over wireless channels. As we shall see, this lower bound (and the underlying coding scheme described in Section~\ref{sec:ach}) can be used to improve the wireless DC system in \cite{li_wireless_2019}.}

\section{Application to wireless distributed computing} \label{sec:application_WDC}

\subsection{The MapReduce System} Consider a  distributed computing (DC) system with $\sf K$ nodes labelled $1,\ldots, \sf K$;  $\sf N$ input files  $W_1,\ldots, W_{\sf{N}}$; and $\sf{Q}$ output functions $h_1, \ldots, h_{\sf Q}$ mapping the input files to the desired computations.  
A \emph{Map-Reduce} System decomposes the functions $h_1,\ldots, h_{\sf{Q}}$   as 
\begin{equation}
h_q( W_1,\ldots, W_{\sf N})= \phi_q( a_{q,1}, \ldots, a_{q,\sf N}), \qquad q \in [\sf{Q}],
\end{equation}
where $\phi_q$ is an appropriate \emph{reduce function} and $a_{q,i}$ is an \emph{intermediate value (IVA)} calculated from input file $W_i$ through an appropriate \emph{map function}:
\begin{equation}
a_{q,i}= \psi_{q,i}(W_i), \qquad i\in[\sf N].
\end{equation} 
For simplicity,  all IVAs are assumed independent and consisting of $\sf{A}$ i.i.d. bits. 

Computations are  performed in  3 phases:

\textbf{Map phase}: A subset of all input files $\mathcal{M}_p\subseteq [\sf N]$ is assigned to each  node $p\in [\sf K]$. Node $p$  computes all IVAs $\{a_{q,i} \colon i \in \mathcal{M}_p, q \in [\sf Q]\}$ associated with these input files.
   
 \textbf{Shuffle phase:} Computations of the $\sf Q$ output functions is assigned to the $\sf K$ nodes, where we denote by $\mathcal{Q}_p\subseteq [\sf Q]$ the output functions assigned to node $p$. 
 
    The $\sf K$ nodes in the system communicate over $\sf T$ uses of a wireless network in a full-duplex mode, where $\sf T$ is a design parameter.  During this communication,  nodes communicate IVAs that they calculated in the Map phase to nodes that are missing these IVAs for the computations of their assigned output functions.  So, node $p\in[\sf  K]$  produces  complex channel inputs of the form
    \begin{equation}\label{eq:comp_encoding}
    X_p^{(\sf  T)}  \triangleq (X_p(1),\ldots, X_p(\sf T))= f_p^{(\sf  T)}\left(\{a_{1, i}, \ldots,  a_{\sf Q, i}\}_{i \in \mathcal M_p}\right),
    \end{equation}  by means of  appropriate encoding function  $f_p^{(\sf  T)}$ satisfying the power constraint \eqref{eq:power}.
    Given the full-duplex nature of the network, Node $p$  also observes  the complex channel outputs
    \begin{equation}
    Y_{p}(t)= \sum_{\ell \in [\sf K]}H_{p,\ell} (t) X_{\ell}(t) + Z_{p}(t), \quad t\in[\sf T],
    \end{equation}
    where  noises $\{Z_{p}(t)\}$ and channel coefficients $\{H_{p,\ell}(t)\}$ are as defined in Section~\ref{sec:channel_model}. 
 
Based on its outputs $Y_{p}^{(\sf T)}\triangleq (Y_p(1),\ldots, Y_p(\sf T))$ and the IVAs $\{a_{q,i} \colon i \in \mathcal{M}_p, \; q \in [\sf Q]\}$  it computed during the Map phase, Node~$p$  decodes the missing IVAs $\{a_{q,i}\colon i \notin \mathcal{M}_p, q \in \mathcal{Q}_p\}$ required to  compute its assigned output functions $\{h_q\}_{q\in \mathcal{Q}_p}$ as:
\begin{equation}\label{eq:comp_decoding}
\hat{a}_{q,i} = g_{q,i}^{(\sf  T)}\left(\{a_{1, i}, \ldots, a_{\sf Q, i}\}_{i \in \mathcal M_p}, Y_{p}^{(\sf T)}\right), \quad i \notin \mathcal{M}_p,\; q \in \mathcal{Q}_p. 
\end{equation}
    
 \textbf{Reduce phase:} Each node applies the reduce functions to the appropriate  IVAs calculated during the Map phase  or decoded in the Shuffle phase.

The performance of the distributed computing system is  measured  in terms of its \emph{computation load}
\begin{equation}\label{eq:r}
   \sf r \triangleq \sum_{p \in [\sf K]} \frac{|\mathcal{M}_p|}{\sf N},
\end{equation}
and the \emph{normalized delivery time (NDT)}
\begin{equation}\label{eq:Delta}
    \sf \Delta =\varliminf_{\sf P \rightarrow \infty} \varliminf_{\sf A \to \infty} \frac{\sf T}{\sf A \cdot  \sf Q\cdot \sf N} \cdot \log \sf P.
\end{equation}

We focus on the \emph{fundamental NDT-computation tradeoff} $\Delta^*(\sf r)$, which is defined as  the  infimum over all values of  $\Delta$ satisfying \eqref{eq:Delta} for some choice of file assignments $\{\mathcal{M}_p\}$, transmission time $\sf T$, and encoding and decoding functions $\{f_p^{(\sf T)}\}$ and $\{g_{q,i}^{(\sf T)}\}$ in \eqref{eq:comp_encoding} and \eqref{eq:comp_decoding}, all depending on $\sf A$ so that the probability  IVA decoding error
\begin{equation}\label{eq:error_computing}
 \textnormal{Pr}\bigg[ \bigcup_{ \substack{p\in[{\sf K}]}}\;  \bigcup_{ q\in\mathcal{Q}_p } \; \bigcup_{i \notin\mathcal{M}_p}  \hat{a}_{q,i}\neq a_{q,i} \bigg] \to 0 \quad \textnormal{as} \quad \sf A \to \infty.
\end{equation} 

\subsection{Results on Normalized Delivery Time} 

\lo{Based on the lower bound in \eqref{eq:PUDoF} we obtain:}
\begin{theorem}\label{thm:DC}Assume $\sf N$ and $\sf Q$ are both multiples of $\sf K$.
If $\sf N$ is large enough, the fundamental NDT-computation tradeoff of the full-duplex wireless DC system is upper bounded as
\begin{IEEEeqnarray}{rCl}\label{eq:DU}
\lefteqn{\Delta^*(\sf  r) \leq} \nonumber \\
&& \textnormal{lowc}\left( (\mathsf{K}, 0)\cup \left\{ \left(\sf r, \, \frac{1-\sf r/{\sf K} }{\textnormal{SDoF}_{\textnormal{Lb}}} \right)\colon \sf 1 \leq \sf r <\sf K\textnormal{ and }\sf r | \sf  K\right\}\right),\IEEEeqnarraynumspace
\end{IEEEeqnarray}
where lowc$(\cdot)$ denotes the lower-convex envelope, $\textnormal{SDoF}_{\textnormal{Lb}}$ is defined in Eq. \eqref{eqn:dof_lower_bound}, and  $\sf r | \sf K$ indicates that $\sf r$ divides $\sf K$.
\end{theorem}
\begin{IEEEproof}
We  prove the result for  integer  values of $\sf  r \in[\sf  K]$ that divide $\sf K$. The final result  follows  by time- and memory-sharing arguments when $\sf N$ is sufficiently large. 

We reuse the  group definition in $\TS{k}$ in \eqref{eq:Group}.

 \textbf{Map phase:}  Choose the same file assignment for all nodes in group $\TS{k}$: 
 \begin{IEEEeqnarray}{rCl}\label{eq:sameINPUT}
 \mathcal{M}_p& = & \tilde{\mathcal{M}}_k\triangleq \left\{ (k-1) \frac{\sf  r \sf N}{\sf K} + 1, \ldots, k \frac{ \sf r \sf N}{\sf K} \right\},  \nonumber\\
&& \hspace{3.5cm}  p\in \TS{k},\; k\in[\tilde{\sf K}],
 \end{IEEEeqnarray}
 This file assignment satisfies the communication load $\sf r$ in  \eqref{eq:r}.

 \textbf{Shuffle phase:} We assume the output function assignment is given as: 
\begin{equation} 
\mathcal{Q}_p \triangleq \{(p-1)\sf Q /\sf K +1,\ldots, p \sf  Q /\sf K\}, \quad p \in [\sf  K]. 
\end{equation}

\lo{Further, choose a sequence (in $\sf P>0$) of rates $\sf R(\sf P)>0$ such that 
\begin{equation}
\varlimsup_{P \to \infty} \frac{\sf R(\sf P)}{\log \sf P} =\frac{ \textnormal{SDoF}_{\textnormal{Lb}}}{\sf K(\sf K/\sf r-1)}
\end{equation}
and such that  for each $\sf P$ the symmetric rate-tuple $(\sf R_{p,k}=\sf R(\sf P), k \in [\tilde{\sf K}], p \in [\sf K]\backslash \TS{k})$ lies
inside the capacity region  $\mathcal{C}(\sf P)$ for the setup in Section~\ref{sec:channel_model}. Fix a power $\sf P$  and consider a sequence (in $\sf T'$) of coding schemes $\{f_p^{(\sf T')}\}_{\sf T'}$ and $\{g_{p,k}^{(\sf T')}\}_{\sf T'}$ for the chosen rate-tuple  such that  $p^{(\sf T')}(\textnormal{error})$ in \eqref{eq:error_prob} tends to $0$ as $\sf T' \rightarrow \infty$. By 
 Theorem~\ref{thm:dof_bounds} and Remark~\ref{rem:PUDoF}, all the mentioned sequences exist. 
}

The shuffle phase is split into rounds, where 
in each round, each group of nodes $\TS{k}$  communicates a different IVA $a_{\nu,i}$  to each node $\ell\in [\sf K]\backslash \RS{k}$, for chosen $\nu \in \mathcal{Q}_{\ell}$ and $i \in \tilde{\mathcal{M}}_{k}$.
To send all missing IVAs, $\Phi \triangleq |\mathcal{Q}_{1}|\cdot |\tilde{\mathcal{M}}_{1}|=(\sf Q/ \sf K)\cdot  (\sf N \sf r /\sf K)$ rounds are necessary. 

 \lo{Any  node $p\in \TS{k}$  uses the chosen encoding function $f_p^{(\sf T')}$ to send the IVAs in a given round, 
for a blocklength $\sf T'$ satisfying}
\sh{Choose a blocklength $\sf T'$ and each node chooses an encoding function $\{f_p^{(\sf T')}\}_{T'}$ achieving symmetric rates ($\sf R_{p,k}=\sf R(\sf P), k\in [\tilde{\sf K}], p \in [\sf K]\backslash\TS{k}$) according to the setup of Section~\ref{sec:channel_model} to encode the IVAs transmitted in a round. The rate $\sf R(\sf P)$ satisfies   $\varlimsup_{\sf P \rightarrow \infty} \frac{\sf R(\sf P)}{\log \sf P}= \frac{ \textnormal{SDoF}_{\textnormal{Lb}}}{\sf K(\sf K/\sf r-1)}$ and $\sf T'$ is chosen such that}
\begin{equation}\label{eq:Tprime}
\frac{\sf A}{\sf T'} < \sf R(\sf P).
\end{equation}
\sh{ Since   all nodes in a group $\RS{k}$ compute the same IVAs in the Map phase,  they can  compute each others' inputs. With its channel outputs in a  round, each Node $p\in \mathcal{T}_k$  thus    forms: 
\begin{equation}
\tilde{Y}_{p}(t) \triangleq Y_p(t) - \sum_{\ell \in \RS{k}} H_{p,\ell}(t) X_{\ell}(t), \quad  p \in \TS{k}, \; t \in[\sf T'],
\end{equation}
and  it applies the decoding functions $\{g_{p,k}^{(\sf T')}\colon k \in  [\tilde{\sf K}] \backslash \lceil p/ \sf r\rceil\}$ corresponding to $\{f_{p}^{(\sf T')}\}$ to the sequence $\tilde{Y}_{p}(1), \ldots, \tilde{Y}_{p}(\sf T')$  to reconstruct the IVAs sent to it in this round.}

\lo{Notice that all nodes in a  group $\RS{k}$ compute the same IVAs in the Map phase, and they can thus compute each others' inputs. Therefore, after receiving its channel outputs $Y_p^{(\sf T')}$ in a given round, any Node $p\in \mathcal{T}_k$ first uses  the IVAs it calculated during the Map phase  to reconstruct and mitigate the signals sent by Txs in the same group $\TS{k}$:
\begin{equation}
\tilde{Y}_{p}(t) \triangleq Y_p(t) - \sum_{\ell \in \RS{k}} H_{p,\ell}(t) X_{\ell}(t), \quad  p \in \TS{k}, \; t \in[\sf T'].
\end{equation}
Then, it applies the chosen decoding functions $\{g_{p,k}^{(\sf T')}\colon k \in  [\tilde{\sf K}] \backslash \lceil p/ \sf r\rceil\}$ to reconstruct the IVAs sent to it in this round from all Tx-groups except for Tx-group $\lceil p / \sf r \rceil$.}


\textbf{Analysis:} 
By our choice of the coding scheme and (\ref{eq:Tprime}), the probability of error in \eqref{eq:error_computing} tends to 0 as $\sf T' \to \infty$. By (\ref{eq:PUDoF}), (\ref{eq:Tprime}) and since the total length of the shuffle phase is $\sf T\triangleq \Phi\sf T'$, the NDT of our scheme is:
\begin{IEEEeqnarray}{rCl}
\nonumber  \varliminf_{\sf P \to \infty} \varliminf_{\sf A \to \infty} \frac{\sf T\log \sf P}{\sf A \cdot  \sf Q\cdot \sf N} &= & \varliminf_{\sf P \to \infty}  \varliminf_{\sf A \to \infty}  \frac{\Phi\sf T'\log \sf P}{\sf A \cdot  \sf Q\cdot \sf N}\\
  &\geq &  \varliminf_{\sf P \to \infty}\frac{\sf r }{\sf K^2} \frac{\log \sf P}{\sf R(\sf P) } 
= \frac{1- \frac{\sf r }{\sf K}}{\textnormal{SDoF}_{\textnormal{Lb}} }.
\end{IEEEeqnarray}
 This proves the desired achievability result.
\end{IEEEproof}

The one-shot scheme in \cite{li_wireless_2019}, which applies  zero-forcing and side information cancellation,  achieves the   upper bound
\begin{IEEEeqnarray}{rCl}\label{eq:DU2}
 \Delta^*(\sf  r) \leq  \textnormal{lowc}\left(\left\{ \left( \sf r, \; \frac{1-{\sf r/\sf K}}{\min(\sf K, 2\sf r)} \right) \colon 1 \leq \sf r \leq \sf K \right\} \right).\IEEEeqnarraynumspace
\end{IEEEeqnarray}
For fixed $\sf K$ and for $\sf  r$ a value that divides $\sf K$ but neither equals $\sf K/2$ nor $\sf K/3$, our new upper bound in \eqref{eq:DU} is strictly better (lower) than the upper bound in \eqref{eq:DU2}. If $\sf K$ is even, the two bounds  coincide on the interval $\sf r  \in [\sf K/2,\sf K]$, where they are given by the straight line $(1-\sf r /\sf K)/\sf K$. If  $\sf K$  is a multiple of $3$,  the two bounds also coincide for $\sf r =\sf K/3$, where they are given by $(1- \sf r/ \sf K)/\sf r$.  For other values of $\sf r$, the bound in \eqref{eq:DU2} can be smaller. An improved upper bound on $\Delta^*(\sf  r)$ is thus obtained by combining the two upper bounds, which results in the lower-convex envelope of the union of the sets  in \eqref{eq:DU} and \eqref{eq:DU2}.

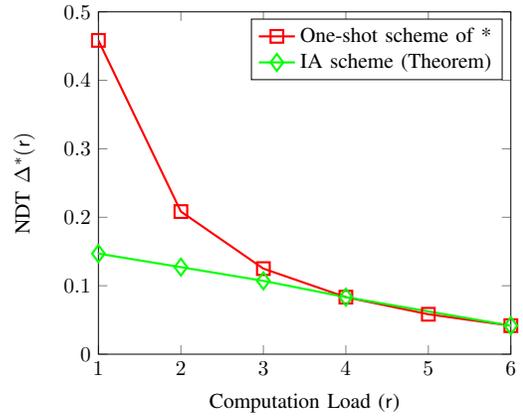
\begin{figure}[h!]

	\centering
	\scalebox{0.8}{
	\begin{tikzpicture}[spy using outlines=
	{circle, lens={xscale=4, yscale=4}, connect spies}]
		\begin{axis}[
			xmin=1, xmax=6, xlabel={Computation Load ($\sf r$)},
			ymin=0, ymax=0.5, ylabel={NDT $\Delta^*(\sf r)$},
			]
			\addplot [color=red, line width=1, mark=square, mark size=3] table [x index=0, y index=1]{figures/res_figure_data/ONE_SHOT.txt};
			\addlegendentry{One-shot scheme of *}
			\addplot [color=green, line width=1, mark=diamond, mark size=4] table [x index=0, y index=1]{figures/res_figure_data/IA.txt};
			\addlegendentry{IA scheme (Theorem)}
		\end{axis}
	\end{tikzpicture}
	}
	\caption{Upper bounds on $\Delta^*(\sf r)$ for the  one-shot scheme in * and  our IA scheme when $\sf K = 12$.}
	\label{fig:comparison}
\end{figure}

In Fig. \ref{fig:comparison}, we  numerically compare the bounds in \eqref{eq:DU} and \eqref{eq:DU2} for $\sf K=12$. We observe that on the interval $\sf r \in [0,4]$ the bound in \eqref{eq:DU} performs better and on the interval $\sf r \in [4,6]$ the bound \eqref{eq:DU2} performs better because \eqref{eq:DU} is simply given by a straight line as $\sf r=5$ does not divide $12$. On the interval $\sf r \in[6,12]$ both bounds perform equally-well as explained in the previous paragraph.

%
%
%

\section{Proof of the SDoF Lower Bound in Theorem~\ref{thm:dof_bounds}}\label{sec:ach}

\sh{For $\tilde{\sf K}\in\{2,3\}$ the proof follows easily by ignoring all Txs, Rxs, and messages outside groups $\TS{1}$ and $\TS{2}$. For details see \cite[Section~V.A]{arxiv2022}. We thus focus on the case $\tilde{\sf K}=4$.}
\lo{\subsection{Proof for $\tilde{\sf K}\in\{2,3\}$}
Choose Tx/Rx-groups $\TS{1}$ and $\TS{2}$,  and ignore all Txs, Rxs, and messages in the other groups. This reduces the network into two non-interfering $\sf r$-user broadcast channels, one from Tx-group $\TS{1}$ to Rx-group $\RS{2}$ and the other  from Tx-group $\TS{2}$ to Rx-group $\RS{1}$, where SDoF $\sf r$ is achievable on each of them.

\subsection{Proof for $\tilde{\sf K}\geq 4$}
}

\subsubsection{Coding Scheme}
We fix a parameter $\eta \in \mathbb{Z}^+$, define 
\begin{IEEEeqnarray}{rCl}
\Gamma  \triangleq  \sf K (\sf K-2), 
\end{IEEEeqnarray}
\lo{and choose 
\begin{IEEEeqnarray}{rCl}
\sf T =  \eta^{\Gamma}(\tilde{\sf K}-2)+(\eta+1)^{\Gamma} (\tilde{\sf K}-1).
\end{IEEEeqnarray}}
\sh{and choose $\sf T =  \eta^{\Gamma}(\tilde{\sf K}-2)+(\eta+1)^{\Gamma} (\tilde{\sf K}-1).$}
Each message $\{a_{p,k}\}$ for $k \in[\tilde{\sf K}], \; p \in [\sf K]\backslash \TS{k}$---but not  messages $\{a_{p, \tilde{\sf K}}\}_{p\in \TS{1}}$ which are not transmitted in our coding scheme---is encoded using a circularly symmetric Gaussian codebook of average power $\sf P /(\sf K- \sf r)$ and  codeword  length $\eta^\Gamma$. Each codeword is sent over a block of $\sf T$ consecutive channel uses.
\sh{More precisely, let $\{\vect{b}_{p,k}\}$ be the $\eta^\Gamma$-length codeword for message $a_{p,k}$. For  $j\in[ \tilde{\sf K}]$, $k \in [\tilde{\sf K}]\backslash \{j\}$ and $(j,k) \neq (1,\tilde{\sf K})$ define 
 $\tilde{\vect{b}}_{j,k} \triangleq \big( \vect{b}_{(j-1)\sf r+1, k}^\intercal,  \vect{b}_{(j-1)\sf r+2, k}^\intercal, \cdots, \vect{b}_{j \cdot \sf r, k}^\intercal	\big)^\intercal$ where the superscript $^\intercal$ stands for the transpose operator. Form also the $\sf T$-length vector of channel  inputs
$\vect{X}_q  \triangleq (X_{q}(1),\ldots,X_{q}(\sf T) )^\intercal$ for each Tx~$q$, and  
$\tilde{\vect{X}}^{(k)} \triangleq \big(\vect{X}_{(k-1) \sf r+1}^\intercal, \cdots, \vect{X}_{k \sf r}^\intercal  \big)^\intercal$ for each Tx-group $k$.} 
\lo{More precisely, let $\vect{b}_{p,k}$ denote the $\eta^\Gamma$-length codeword symbol for message $a_{p,k}$ and define
for each $j\in[ \tilde{\sf K}]$, $k \in [\tilde{\sf K}]\backslash \{j\}$ and $(j,k) \neq (1,\tilde{\sf K})$ the vector
\begin{equation}
\tilde{\vect{b}}_{j,k} \triangleq \left( \vect{b}_{(j-1)\sf r+1, k}^\intercal,  \vect{b}_{(j-1)\sf r+2, k}^\intercal, \cdots, \vect{b}_{j \cdot \sf r, k}^\intercal	\right)^\intercal.
\end{equation}
Group the channel inputs and outputs into the vectors
\begin{IEEEeqnarray}{rCl}
	\vect{X}_q & \triangleq& (X_{q}(1),\ldots,X_{q}(\sf T) )^\intercal, \quad q\in[\sf K],\\
	\vect{Y}_p &\triangleq &(Y_{p}(1),\ldots,Y_{p}(\sf T) )^\intercal, \quad p \in[\sf K],
\end{IEEEeqnarray}
and for each Tx-group $k \in [\tilde{\sf K}]$: 
\begin{equation}
\tilde{\vect{X}}^{(k)} \triangleq \begin{pmatrix}\vect{X}_{(k-1)\cdot \sf r+1} \\ \vdots\\ \vect{X}_{k\cdot \sf r},
\end{pmatrix}
\end{equation}
and Rx-group $j \in [\tilde{\sf K}]$: 
\begin{equation}
\tilde{\vect{Y}}^{(j)} \triangleq \begin{pmatrix}\vect{Y}_{(j-1) \cdot \sf r+1} \\ \vdots \\\vect{Y}_{j \cdot \sf r} \end{pmatrix} = \sum_{k \neq j} \tilde{\mat{H}}^{(j,k)} \tilde{\vect{X}}^{(k)}  + \tilde{\vect{Z}}^{(j)},
\end{equation}
where $\tilde{\vect{Z}}^{(j)}$ is the corresponding Gaussian vector and $\tilde{\mat{H}}^{(j,k)}$ is the $\sf r\sf T \times \sf r\sf T$ channel matrix 
\begin{equation}\label{eq:bigH}
 \tilde{\mat{H}}^{(j,k)} \triangleq \begin{pmatrix} 
 \mat{H}_{(j-1)\cdot \sf r +1, (k-1)\cdot \sf r +1} &\cdots  & \mat{H}_{(j-1)\cdot \sf r +1, k\cdot  \sf r } \\ \vdots & & \vdots  \\
 \mat{H}_{j\cdot \sf r, (k-1)\cdot \sf r +1} & \cdots & \mat{H}_{j\cdot \sf r , k\cdot \sf r } 
 \end{pmatrix}
\end{equation}
and
\begin{equation}
 \mat{H}_{p,q} \triangleq \text{diag}(\left[H_{p,q}(1), H_{p,q}(2) \cdots H_{p,q}(\sf T)\right]).
\end{equation}}
Tx-groups 
form their inputs as:
\begin{IEEEeqnarray}{rCl}
\tilde{\vect{X}}^{(1)} &=& \sum_{i=2}^{\tilde{\sf K}} \tilde{\mat{V}}^{(i,1)} \tilde{\mat{U}}_i \tilde{\vect{b}}_{i,1}, \label{eq:X1}\\
\tilde{\vect{X}}^{(k)}& =& \sum_{i\in[\tilde{\sf K}]\backslash \{1,k\}}  \tilde{\mat{V}}^{(i,k)}  \tilde{\mat{U}}_i \tilde{\vect{b}}_{i,k} 
+ \tilde{\mat{V}}^{(1,k)}  \tilde{\mat{U}}_k  \tilde{\vect{b}}_{1,k}, \nonumber \\[-2ex]
&&\hspace{3.4cm}\quad k\in[\tilde{\sf K}-1]\backslash\{1\}, \IEEEeqnarraynumspace \label{eq:X2}\\
\tilde{\vect{X}}^{(\tilde{\sf K})} &=& \sum_{i=\{2,\ldots, \tilde{\sf K}-1\}}   \tilde{\mat{V}}^{(i,\tilde{\sf K})} \tilde{\mat{U}}_i \tilde{\vect{b}}_{i,\tilde{\sf K }} ,\label{eq:XK}
\end{IEEEeqnarray}
where 
\begin{equation}
\tilde{\mat{U}}_i \triangleq 
\mat{Id}_{\sf r} \otimes \mat{U}_i, \quad i \in \{2,3,\cdots, \tilde{\sf K}\},
\end{equation}
and 
 matrices $\{\mat{U}_i\}$ and $\{\mat{V}^{(i,k)}\}$ are described shortly.

Notice  that for $i \in \{2,3,\cdots, \tilde{\sf K}\}$, messages $\left\{ \tilde{\vect{a}}_{i,k} \right\}_{k \in [\tilde{\sf K}]\backslash\{i\}}$ and $\tilde{\vect{a}}_{1,i}$ are multiplied by the same precoding matrix $\tilde{\mat{U}}_i$.

\subsubsection{Zero-forcing Matrices $\{\tilde{\mat{V}}^{(i,k)}\}$}
For each $i,k \in [\tilde{\sf K}]$ with $i \neq k$,  construct the $\sf T \times \sf T$ diagonal matrices $\mat{S}^{(i,k)}_1, \cdots, \mat{S}^{^{(i,k)}}_r$ by picking the real and imaginary parts of all non-zero entries i.i.d. according to a continuous distribution over $[-\sf H_{\max}, \sf H_{\max}]$ and form the diagonal matrix 
\begin{IEEEeqnarray}{rCl}
\mat{S}^{(i,k)} \triangleq
\sh{\textnormal{diag}\big(  \mat{S}^{(i,k)}_1 , \ldots,   \mat{S}^{^{(i,k)}}_{\sf r}\big)}
\lo{
\begin{pmatrix} \mat{S}^{(i,k)}_1 & \mat{0} &    \mat{0}\\ \mat{0}&  \ddots & \mat{0} \\
\mat{0} & \cdots &  \mat{S}^{^{(i,k)}}_{\sf r}\end{pmatrix}  }. \IEEEeqnarraynumspace
\end{IEEEeqnarray}
\sh{Define the $\sf r\sf T \times \sf r\sf T$ channel matrix from Tx-group $k$ to Rx-group $j$:
\begin{equation}
 \tilde{\mat{H}}^{(j,k)} \triangleq \begin{pmatrix} 
 \mat{H}_{(j-1)\cdot \sf r +1, (k-1)\cdot \sf r +1} &\cdots  & \mat{H}_{(j-1)\cdot \sf r +1, k\cdot  \sf r } \\ \vdots & & \vdots  \\
 \mat{H}_{j\cdot \sf r, (k-1)\cdot \sf r +1} & \cdots & \mat{H}_{j\cdot \sf r , k\cdot \sf r } 
 \end{pmatrix}
\end{equation}
where $\mat{H}_{p,q} \triangleq \text{diag}(\left[H_{p,q}(1), H_{p,q}(2), \ldots,  H_{p,q}(\sf T)\right])$.}

Choose the precoding matrices as:\footnote{We assume that all  matrices $\{\tilde{\mat{H}}^{(i,k)}\}$ are invertible, which happens with probability 1. Otherwise, Txs and Rxs immediately declare an error in the communication. \lo{This probability-0 event,  does not change the error probability of the system.}}
\begin{IEEEeqnarray}{rCl}
 \tilde{\mat{V}}^{(i,k)} = 	 \left( \tilde{\mat{H}}^{(i,k)} \right)^{-1} \mat{S}^{(i,k)},    \qquad  i,k \in [\tilde{\sf K}], \; i\neq k,
\end{IEEEeqnarray}
so that all information sent to any Rx in group $\TS{j}$ is zero-forced at all other Rxs in the same group $\TS{j}$.
Defining for each triple $(i,j,k)\in [\tilde{\sf K}]^3$ with $i\neq j, j \neq k ,k \neq i $ the ``generalized" channel matrix 
\begin{IEEEeqnarray}{rCl}
\tilde{\mat{G}}^{(i,k)}_{j} &=& 
\begin{pmatrix} 
 \mat{G}_{(j-1)\cdot \sf r +1}^{((i-1)\cdot \sf r +1, k)} &\cdots  & \mat{G}_{(j-1)\cdot \sf r +1}^{(i\cdot \sf r, k) } 
 \\ \vdots & & \vdots  \\
 \mat{G}_{j\cdot \sf r}^{ ((i-1)\cdot \sf r +1, k)} & \cdots & \mat{G}_{(j\cdot \sf r})^{i\cdot \sf r , k} 
 \end{pmatrix} 
 \label{eq:def_matrix_G}\\
&\triangleq& \tilde{\mat{H}}^{(j,k)} \cdot \tilde{\mat{V}}^{(i,k)} =  \tilde{\mat{H}}^{(j,k)}  \left(  \tilde{\mat{H}}^{(i,k)}  \right)^{-1} \mat{S}^{(i,k)}, \IEEEeqnarraynumspace \label{eq:Gexp}
\end{IEEEeqnarray}
allows to write the  signals at the various Rx-groups as: 
\begin{IEEEeqnarray}{rCl}\label{eq:Y1}
\tilde{\vect{Y}}^{(1)} 
&= &\underbrace{\sum_{k=2}^{\tilde{\sf K}-1}\mat{S}^{(1,k)}    \tilde{\mat{U}}_k   \tilde{\vect{b}}_{1,k}}_{\text{desired signal}}+\sum_{i=2}^{\tilde{\sf K}}  \sum_{k\notin \{1,i\}} \tilde{\mat{G}}_1^{(i,k)}   \tilde{\mat{U}}_i   \tilde{\vect{b}}_{i,k} + \tilde{\vect{Z}}^{(1)},\nonumber\\[-2ex]\\
\label{eq:Yj}
\tilde{\vect{Y}}^{(j)} 
&=  &\underbrace{\sum_{k \neq j} \mat{S}^{(j,k)}    \tilde{\mat{U}}_j   \tilde{\vect{b}}_{j,k}}_{\text{desired signal}}
+ \sum_{i \notin \{1, j\}}  \sum_{k \notin \{i,j\}}  \tilde{\mat{G}}_j^{(i,k)}   \tilde{\mat{U}}_i   \tilde{\vect{b}}_{i,k}\nonumber  \\
&& + \sum_{k \notin \{1, j, \tilde{K}\}} \tilde{\mat{G}}_j^{(1,k)} \tilde{\mat{U}}_k  \tilde{\vect{b}}_{1,k} + \tilde{\vect{Z}}^{(j)}, \quad j\in[ \tilde{\sf{K}}]\backslash\{1\}.\nonumber\\[-2ex]
\end{IEEEeqnarray}
\sh{Here $\tilde{\vect{Z}}^{(j)}$ are the corresponding Gaussian noise vectors observed at Rx-group~$j$, for $j\in[\tilde{\sf K}]$.} \lo{The third sum in \eqref{eq:Yj} has $\tilde{K}-2$ terms when $j = \tilde{K}$ but only $\tilde{K}-3$ terms otherwise.}

\subsubsection{IA Matrices $\{\mat{U}_{i}\}$} 
Inspired by the IA scheme in \cite{jafar_degrees_2008}, we choose each $\sf T \times \eta^\Gamma$ precoding matrix $\mat{U}_i$ so that its column-span includes all power products (with powers from 1 to $\eta$) of the ``generalized" channel matrices $\mat{G}_{p'}^{(p,k)}$ 
 that premultiply $\mat{U}_i$ in \eqref{eq:Y1} and \eqref{eq:Yj}. That means for $i\in[ \tilde{\sf K}]\backslash\{1\}$:
\begin{IEEEeqnarray}{rCl}\label{eq:U_i}
\mat{U}_{i}= \left[ \prod_{\mat{G} \in {\mathcal{G}}_{i} }
\mat{G}^{{\alpha_{i,\mat{G}}}}\cdot  \boldsymbol{\Xi}_{i}\colon  \, \; 
\forall \boldsymbol{\alpha}_i \in [\eta]^{\Gamma\cdot \sf r^2}\right], \quad 
\end{IEEEeqnarray}
where 
$\{\boldsymbol{\Xi}_i\}_{i=2}^{\sf K}$ are i.i.d. random vectors independent of all channel matrices, noises, and messages, and 
\begin{IEEEeqnarray}{rCl}
\label{eq:set_G}
\mathcal{G}_i& \triangleq& \left\{  \mat{G}_{p'}^{(p,k)} \colon p \in \TS{i},  \;  k \in[\tilde{\sf K}] \backslash \TS{i}, \; p' \in [\sf{K}]\backslash  (\TS{i} \cup \TS{k}) \right\}\nonumber  \\
&  &\cup \left \{  \mat{G}_{p}^{(p',i)} \colon p \in \TS{1}, \;  p' \in[\sf K]\backslash \{ \TS{i} \cup \TS{1}\}  \right\},
\end{IEEEeqnarray}
and $\boldsymbol{\alpha}_i \triangleq (\alpha_{i, \mat{G}}\colon \quad \mat{G} \in {\mathcal{G}}_{i}).$ 

\subsubsection{Analysis of Signal-and-Interference Subspaces}
Since the column-span of $\mat{U}_i$ contains all power products  of powers 1 to $\eta$ of the modified channel matrices $\mat{G} \in \mathcal{G}_i$ that premultiply $\mat{U}_i$ in \eqref{eq:Y1} and \eqref{eq:Yj}, the product of any of these matrices with $\mat{U}_i$ is included in the column-space of the $\sf T\times  \eta^\Gamma$-matrix
\begin{IEEEeqnarray}{rCl}\label{eq:W_i}
\mat{W}_{i} =  \left[ \prod_{\mat{G} \in {\mathcal{G}}_{i} }
\mat{G}^{{\alpha_{i,\mat{G}}}} \cdot \boldsymbol{\Xi}_{i}\colon  \; \, 
\forall \boldsymbol{\alpha}_i \in [\eta+1]^{\Gamma\cdot r^2}\right]&,
\nonumber\\
\textnormal{ for } i\in[ \tilde{\sf K}]\backslash\{1\}&,
\end{IEEEeqnarray}
where notice that $|\mathcal{G}_i|=\Gamma$.
Formally, for each $i \in \{2,3, \cdots, \tilde{\sf K}\}$ and $\mat{G} \in \mathcal{G}_i$, we have $\text{span}(\mat{G}\cdot \mat{U}_i) \subseteq \text{span}(\mat{W}_i)$. 
As a consequence, the signal and interference space at a Rx~$p\in\TS{j}$, for $j\in\{2, \ldots, \tilde{\sf K}\}$, is represented by the 
matrix: 
\begin{IEEEeqnarray}{rCl}\label{eq:signal_interference2}
\boldsymbol{\Lambda}_p \triangleq
 &&   \big[ \underbrace{\mat{D}_p,}_{\textnormal{signal space}} \underbrace{ {\mat{W}}_{2}, \;\cdots,\; {\mat{W}}_{j-1},\; {\mat{W}}_{j+1},\;\cdots,\; {\mat{W}}_{\tilde{K}}}_{\textnormal{interference space}} \big] .  \IEEEeqnarraynumspace
\end{IEEEeqnarray}
with the signal subspaces given by the $\sf T \times (\tilde{\sf K}-1)\eta^{\Gamma}$-matrices
\begin{IEEEeqnarray}{rCl}
    \mat{D}_p  \triangleq \left[\mat{S}^{(j,k)}_{p \textnormal{ mod } \sf r} \cdot \mat{U}_j \right]_{k \in [\tilde{\sf K}]\backslash\{j\}}, \quad p\in \TS{j}.
\end{IEEEeqnarray}
For a Rx $p$ in the first group $\TS{1}$, the signal and interference spaces are represented by the $\sf T \times \sf T$-matrix:
\begin{IEEEeqnarray}{l}\label{eq:signal_interference1}
\boldsymbol{\Lambda}_p=
\big[ \underbrace{\mat{D}_{p,2}, \; \cdots,  \; \mat{D}_{p,\tilde{\sf K}-1},}_{\textnormal{signal space}}  \;  \underbrace{{\mat{W}}_{2},\; {\mat{W}}_{3} , \;\cdots, \;{\mat{W}}_{\tilde{\sf K}}}_{\textnormal{interference space}} \big], \IEEEeqnarraynumspace
\end{IEEEeqnarray}
where the signal subspace is given by the $\sf T \times \eta^{\Gamma}$-matrices
\begin{IEEEeqnarray*}{rCl}
    \mat{D}_{p,k} \triangleq \mat{S}^{(1,k)}_p \cdot \mat{U}_k, \quad  k \in \{2,...,\tilde{\sf K}-1\} ,\quad  p \in \TS{1}.
\end{IEEEeqnarray*}

We shall prove that  all matrices $\{\mat{\Lambda}_p\}$ are of full column rank. This proves  that the desired signals  intended for Rx $p$ can be separated from each other and from the 
interference space at this Rx. In the limits $\eta \to\infty$ (and thus $\sf T \to \infty$) and $\sf P\to\infty$, this establishes an DoF of  
$\lim_{\eta \to \infty} \frac{( \tilde{\sf K}-1)\eta^{\Gamma}}{\sf T} = \frac{ \tilde{\sf K}-1 }{ 2 \tilde{\sf K}-3}$ at Rxs $p \in [\sf K]\backslash\TS{1}$ and an DoF of 
  $\frac{\tilde{\sf K}-2 }{ 2 \tilde{\sf K}-3}$ for Rxs $p \in \TS{1}$. 
  \lo{The SDoF is therefore given by
\begin{IEEEeqnarray}{rCl}
	\textnormal{SDoF} &=& \sf r \cdot \left(\frac{ \tilde{\sf K}-1 }{ 2 \tilde{\sf K}-3} \cdot (\tilde{\sf K}-1)+ \frac{ \tilde{\sf K}-2 }{ 2 \tilde{\sf K}-3}\right) \\
	&=& \frac{\sf K\cdot (\sf K - \sf r) -\sf r^2}{2\sf K - 3\sf r} = \textnormal{SDoF}_\textnormal{Lb},
\end{IEEEeqnarray}
 }
\sh{The SDoF of the entire system is thus given by $\textnormal{SDoF}_{\textnormal{Lb}}$,} which  establishes the desired achievability result.

Notice that each matrix $\mat{\Lambda}_p$, for $p\in [\sf K]$, is of the form of the matrix $\mat{\Lambda}$ in Lemma \ref{lma:diag_fullrank} at the end of this section. Defining the matrices  $\{\hat{\mat{U}}_i\}, \{\hat{\mat{W}_i}\}$, $\{\hat{\mat{D}}_p\}$ and $\{\hat{\mat{D}}_{p,k}\}$ in the same way as  $\{{\mat{U}}_i\}, \{{\mat{W}_i}\}$, $\{{\mat{D}}_p\}$, and $\{{\mat{D}}_{p,k}\}$ but with $\boldsymbol{\Xi}_i$ replaced by the all-one vector $\boldsymbol{1}$, it suffices to show that 
with probability 1 all square submatrices of the  following matrices  (which play the roles of $\{\mat{B}_i\}$ when applying Lemma~\ref{lma:diag_fullrank})  are full rank:
\begin{IEEEeqnarray}{rCl}\label{eq:toexamine}
\{\hat{\mat{D}}_p\}_{p \in [\tilde{\sf K}]\backslash \TS{1}}, \quad \{\hat{\mat{W}}_{j}\}_{j=2}^{\tilde{\sf K}},\quad \left\{  \left[ \hat{\mat{D}}_{p,j}, \; \hat{\mat{W}}_{j} \right]\right\}_{\substack{p\in \TS{1}\\ j\in\{2,\ldots, \tilde{\sf K}\}}} \!\!. \IEEEeqnarraynumspace
\end{IEEEeqnarray}
For matrix $\hat{\mat{D}}_p$, $p\in \TS{2}$, this proof is provided in \lo{Appendix~\ref{app:Pr}}\sh{\cite[Appendix~B]{arxiv2022}}. For the other matrices the proof is similar.

\begin{lemma} \label{lma:diag_fullrank}
Consider positive integers $n_1, n_2, \cdots, n_{\tilde{\sf K}}$ summing to $C \triangleq \sum_{i=1}^{\tilde{\sf K}} n_i \leq \sf T$, and  for each $i \in [\tilde{\sf K}]$ and $k \in [n_i]$ a diagonal $\sf T\times \sf T$ matrix  $\mat{B}_{i,k} \in \mathbb{C}$ so that  all square sub-matrices of the following matrices  are full rank:
\begin{equation}
    \mat{B}_{i} \triangleq \left[ \mat{B}_{i,1} \cdot \boldsymbol{1}, \mat{B}_{i,2} \cdot \boldsymbol{1}, \cdots, \mat{B}_{i,n_i} \cdot \boldsymbol{1}\right], \quad  i \in [\tilde{\sf K}].
\end{equation}

Let   $\{\boldsymbol{\Xi}_i\}$ be independent $\sf T$-length vectors with i.i.d. entries  from continuous distributions and define the $\sf T\times n_i$-matrices
\begin{equation}
    \mat{A}_i \triangleq \left[ \mat{B}_{i,1} \cdot \boldsymbol{\Xi}_{i}, \mat{B}_{i,2} \cdot \boldsymbol{\Xi}_{i}, \cdots, \mat{B}_{i,n_i} \cdot \boldsymbol{\Xi}_{i} \right], \quad i \in [\tilde{\sf K}].
\end{equation}

Then, the   $\sf T \times C$-matrix $\mat{\Lambda} \triangleq \left[ \vect{A}_1, \vect{A}_2, \cdots, \vect{A}_{\tilde{\sf K}} \right]$ has full column rank with probability 1.
\end{lemma}
\begin{IEEEproof}
\sh{ }
\lo{
We present the proof for the case  that the matrix $\*\Lambda$ is square, i.e., $C = \sf T$. If $\sf T > C$, we take a square $C$-by-$C$ submatrix of $\* \Lambda$ and  perform the same proof steps on the submatrix.

Define
\begin{equation}
    F \left( \boldsymbol{\Xi}_1, \ldots, \boldsymbol{\Xi}_{\tilde{\sf K}} \right) \triangleq \det (\*\Lambda)
\end{equation}
which is a polynomial of $\boldsymbol{\Xi}_1, \boldsymbol{\Xi}_2, \cdots, \boldsymbol{\Xi}_{\tilde{\sf K}}$ as the determinant is a polynomial of the entries of $\mat{\Lambda}$.

For the vectors
\begin{IEEEeqnarray}{rCl}
\boldsymbol{\xi}_i =[\underbrace{0, \cdots 0,}_{ (n_1+\cdots +n_{i-1})\textnormal{ 0s}} \underbrace{1, \cdots 1, }_{n_i \textnormal{ 1s}} \underbrace{0, \cdots 0}_{ (n_{i+1}+\cdots +n_{\tilde{\sf K}})\textnormal{ 0s}}]^\intercal , \quad i\in[ \tilde{\sf {K}}], \IEEEeqnarraynumspace
\end{IEEEeqnarray}
the polynomial evaluates to
\begin{IEEEeqnarray}{rCl}
F\left( \boldsymbol{\xi}_1,  \ldots, \boldsymbol{\xi}_{\tilde{\sf K}} \right) &=& \det 
\begin{pmatrix}
\mat{B}_1' & \mat{0} & \cdots & \mat{0}\\
\mat{0} & \mat{B}_2' & \cdots & \mat{0}\\
\vdots & \vdots & \ddots & \vdots \\
\mat{0} & \mat{0} & \cdots & \mat{B}_{\tilde{\sf K}}'
\end{pmatrix} \\
&=& \prod_{i=1}^{\tilde{\sf K}} \det(\mat{B}_i') \neq 0 \label{eq:F_inequation2}
\end{IEEEeqnarray}
where $\mat{B}'_i$ is the $n_i \times n_i$ square sub-matrix of $\mat{B}_i$ consisting  of its rows $(n_1+\cdots+n_{i-1}+1)$ to $(n_1+\cdots+n_{i-1}+n_i)$. As all square sub-matrices of $\mat{B}_i$ are full rank, any matrix $\mat{B}_i'$ for $i\in[\tilde{\sf K}]$ is also full-rank. This leads to $\det(\mat{B}_i') \neq 0$ for  $i\in[\tilde{\sf K}]$. Consequently \eqref{eq:F_inequation2} holds.

We conclude that $F$ is a non-zero polynomial and thus $F \left( \boldsymbol{\Xi}_1, \ldots, \boldsymbol{\Xi}_{\tilde{\sf K}} \right) $ equals $0$ with probability $0$ because the entries of $\boldsymbol{\Xi}_1, \boldsymbol{\Xi}_2, \cdots, \boldsymbol{\Xi}_{\tilde{\sf K}}$ are drawn independently from continuous distributions.}
\sh{
See \cite{arxiv2022}.}
\end{IEEEproof}

\section{Conclusion}\label{sec:ccl}
We provided new lower and upper bounds on the sum degrees of freedom (SDoF) of a particular partially-connected  $\sf K$-user X-channel where each group of $\sf r$ consecutive Txs cooperates and sends a message to each Rx outside its group. When $\sf K /\sf r \in \{2,3\}$ the bounds coincide and establish the exact SDoF of the system. The proposed lower bound is used to provide an improved normalized delivery time  (NDT) for wireless distributed Map-Reduce systems. 


\section*{Acknowledgement}
This work has been supported by the European Research Council (ERC) under the European Union's Horizon 2020 under grant agreement No 715111, and in part by National Key R\&D Program of China under Grant No 2020YFB1807504 and National Science Foundation of China Key Project under Grant No 61831007.

\appendices
\lo{\section{Proof of the SDoF Upper Bound in Theorem~\ref{thm:dof_bounds}}\label{sec:converse}
The proof follows immediately by summing up the upper bound in the following Lemma \ref{lma:upper_bound_lemma} for the   $\tilde{\sf K} (\tilde{\sf K}-1)$ distinct pairs $(j,k) \in [\tilde{\sf K}]\times [\tilde{\sf K}]$  with $j\neq k$, and then dividing this sum by $2\tilde{\sf K}-3$, because each rate has been counted $2\tilde{\sf K}-3$ times.

\begin{lemma}\label{lma:upper_bound_lemma}
Let  $\left(\sf R_{p,k}(\sf P)\colon k \in [\tilde{\sf K}], p \in [\sf K]\backslash \RS{k}\right)$ be a rate-tuple in $\mathcal{C}(\sf P)$, for each $\sf P>0$. Then, for any $j, k\in[\tilde{K}]$ with $j\neq k$: 
\begin{IEEEeqnarray}{rCl}\label{eqn:thm_1}
\varlimsup_{\sf P \to \infty} \left[ \sum_{p \in \RS{j}} \;\sum_{\ell \in [\tilde{\sf K}]\backslash\{j\}}  \frac{ \sf R_{p, \ell}}{\log \sf P} + \sum_{p \in [\sf K] \backslash (\TS{k} \cup \RS{j})}  \frac{\sf R_{p,k}}{ \log \sf P} \right] \leq \sf r.\IEEEeqnarraynumspace
\end{IEEEeqnarray}
\end{lemma}

\begin{IEEEproof} Fix $\sf P >0$ and any rate tuple  $\left(\sf R_{p,k}(\sf P)\colon k \in [\tilde{\sf K}], p \in [\sf K]\backslash \RS{k}\right)$ in $\mathcal{C}(\sf P)$. Then consider a sequence of encoding and decoding functions $\{f_q^{(\sf T)}\}$ and $\{g_{p,k}^{(\sf T)} \}$ such that $p^{(\sf T)}(\textnormal{error})$ tends to $0$ as $\sf T \to \infty$. 
	
	Fix a blocklength $\sf T$ and indices  $j, k\in[\tilde{\sf K}]$ with $j\neq k$, and define 
\begin{equation}
\mathcal{F}\triangleq [\sf  K] \backslash (\TS{k} \cup \RS{j})
\end{equation} 
Partition the set of messages into the following three sets 
\begin{IEEEeqnarray}{rCl}
    \mathbf{a}_r & \triangleq &\{ a_{p,\ell }\colon {p \in \RS{j}}, \; \ell \in [\tilde{\sf  K}]\backslash\{j\}\} \\
    \mathbf{a}_t &\triangleq &\{a_{p,k} \colon p \in \mathcal{F}\}\\
    \mathbf{a}_c &\triangleq& \{a_{p,\ell} \colon p \notin  \RS{j}, \; \ell \neq k\}.
\end{IEEEeqnarray}
Finally, denote by $\mathcal{H}$ the set of \emph{all} channel coefficients in the system, and  for any subset $\mathcal{S}\subseteq [\sf  K]$ define $\*Y_{\mathcal{S}} \triangleq \left(Y_p^{(T)} \colon p  \in \mathcal{S}\right)$ and $\*Z_{\mathcal{S}} \triangleq \left(Z_p^{(T)} \colon p \in \mathcal{S} \right)$.

Notice now that by the independence of the IVAs, the channel coefficients, and the noise sequences:
\begin{IEEEeqnarray}{rCl}
\lefteqn{  \sum_{p \in \RS{j}} \;\sum_{\ell \in [\tilde{\sf K}]\backslash\{j\}}  \sf T \sf R_{p, \ell} + \sum_{p \in [\sf K] \backslash (\TS{k} \cup \RS{j})}\sf T  \sf R_{p,k} }\qquad \\
&=&H(\mathbf{a}_r, \mathbf{a}_t)\\
 &=& H(\mathbf{a}_t, \mathbf{a}_r|\mathbf{a}_c,\mathcal{H})\\
    &=& I(\mathbf{a}_t, \mathbf{a}_r; \*Y_{\mathcal{R}_{j}}|\mathbf{a}_c,\mathcal{H}) +
    H(\mathbf{a}_t, \mathbf{a}_r| \mathbf{a}_c, \*Y_{\mathcal{R}_{j}}, \mathcal{H})\\
    &=& h(\*Y_{\mathcal{R}_{j}}| \mathbf{a}_c,\mathcal{H}) - h(\*Z_{\mathcal{R}_{j}})\nonumber \\
    &&+ H(\mathbf{a}_r| \mathbf{a}_c, \*Y_{\mathcal{R}_{j}}, \mathcal{H}) 
    + H(\mathbf{a}_t|\mathbf{a}_r, \mathbf{a}_c, \*Y_{\mathcal{R}_{j}}, \mathcal{H}) \\
    & \leq  & \sf T \cdot \sf r   \log \left(1+ \sf P \sf H_{\max}^2 (\sf K-\sf r) \right)\nonumber\\
    && + H(\mathbf{a}_r| \mathbf{a}_c, \*Y_{\mathcal{R}_{j}}, \mathcal{H}) 
    + H(\mathbf{a}_t|\mathbf{a}_r, \mathbf{a}_c, \*Y_{\mathcal{R}_{j}}, \mathcal{H}).  \IEEEeqnarraynumspace \label{eq:p1}
\end{IEEEeqnarray}

If communication  is reliable,  it is possible to reconstruct $\mathbf{a}_r$ from $\*Y_{\RS{j}}$ with probability of error tending to 0 as $\sf T\to \infty$. Therefore, by Fano's inequality
\begin{IEEEeqnarray}{rCl} \label{eqn:converse_part3}
    &H(\mathbf{a}_r| \mathbf{a}_c,\*Y_{\mathcal{R}_{j}}, \mathcal{H}) \leq \sf T \cdot  \epsilon_{\sf T}, 
\end{IEEEeqnarray}
for some sequence $\{\epsilon_{\sf T}\}$  tending to $0$ as $\sf T \rightarrow \infty$. 

To bound the last summand in \eqref{eq:p1}, we further notice that for reliable communication, Fano's inequality also implies 
\begin{equation}
H(\mathbf{a}_t|\mathbf{a}_r, \mathbf{a}_c,\*Y_{\mathcal{R}_{j}}, \*Y_{\mathcal{F} }, \mathcal{H}) \leq \sf T \tilde{\epsilon}_{\sf T},
\end{equation}
for some sequence $\{\tilde{\epsilon}_{\sf T}\}$  tending to 0 as $\sf T \to \infty$. Thus,
\begin{IEEEeqnarray}{rCl} \label{eqn:converse_part4_1}
   \lefteqn{ H(\mathbf{a}_t|\mathbf{a}_r, \*Y_{\mathcal{R}_{j}},\mathcal{H}) }  \nonumber\\
   &    \leq& H(\mathbf{a}_t|\mathbf{a}_r, \*Y_{\mathcal{R}_{j}}, \mathcal{H}) ] - H(\mathbf{a}_t|\mathbf{a}_r, \mathbf{a}_c \*Y_{\mathcal{R}_{j}}, \*Y_{\mathcal{F} }, \mathcal{H}) + \sf T\tilde{\epsilon}_{\sf T} \nonumber\\\\
   &= & I (\mathbf{a}_t; \*Y_{\mathcal{F}}|\mathbf{a}_r, \mathbf{a}_c, \*Y_{\mathcal{R}_{j}}, \mathcal{H}) 
    +\sf T \tilde{\epsilon}_{\sf T} \\
    &   \stackrel{(a)}{=}& h(\tilde{\*Y}_{\mathcal{F}}|\mathbf{a}_r, \mathbf{a}_c,\tilde{\*Y}_{\mathcal{R}_{j}}, \mathcal{H})
    - h(\*Z_{\mathcal{F}}) +\sf T \tilde{\epsilon}_{\sf T}\\
       &\stackrel{(b)}{=} & \mathbb{P}(E=1) \cdot  h (\tilde{\*Y}_{\mathcal{F}}|\mathbf{a}_r, \mathbf{a}_c, \tilde{\*Y}_{\mathcal{R}_{j}}, \mathcal{H}, E=1) \nonumber \\
       & &  +\mathbb{P}(E=0) \cdot  h( \tilde{\*Y}_{\mathcal{F}}|\mathbf{a}_r, \mathbf{a}_c, \tilde{\*Y}_{\mathcal{R}_{j}}, \mathcal{H}, E=0) \nonumber\\
      &&    - h(\*Z_{\mathcal{F}}) 
       +\sf T \tilde{\epsilon}_{\sf T} \\
    &\stackrel{(c)}{=}&h (\tilde{\*Y}_{\mathcal{F}}|\mathbf{a}_r, \mathbf{a}_c, \tilde{\*Y}_{\mathcal{R}_{j}}, \mathcal{H}, E=1)    - h(\*Z_{\mathcal{F}}) 
    + \sf T\tilde{\epsilon}_{\sf T}\IEEEeqnarraynumspace \\   
    & \stackrel{(d)}{\leq } &h( {\*Z}_{\mathcal{F}}' |\mathbf{a}_r, \mathbf{a}_c,\tilde{\*Y}_{\mathcal{R}_{j}}, \mathcal{H}, E=1)
    - h(\*Z_{\mathcal{F}}) + \sf T \tilde{\epsilon}_{\sf T}\\
        & \stackrel{(e)}{\leq} &h({\*Z}_{\mathcal{F}}')
    - h(\*Z_{\mathcal{F}}) +\sf T\cdot  \tilde{\epsilon}_{\sf T}, \label{eqn:converse_part4_2}
\end{IEEEeqnarray}
where 
\begin{itemize}
\item  in $(a)$ we defined for any  $\mathcal{S}\subseteq [\sf K]$ the tuple  $ \tilde{\*Y}_{\mathcal{S}} \triangleq \big( \tilde{Y}_p^{(T)} \colon p \in \mathcal{S}\big)$ with $ \tilde{Y}_p^{(T)}\triangleq\big (\tilde{Y}_p(1),\ldots, \tilde{Y}_p(\sf T)\big)$ and 
\begin{equation}
    \tilde{Y}_p(t) \triangleq \sum_{\ell \in \TS{k}} H_{p,\ell}(t)  {X}_{\ell}(t)+ Z_p(t), \quad  p \in \mathcal{S};
\end{equation}
\item in $(b)$ we defined he random variable $E$ equal to  $1$ if each input $X_q(t)$, for $q\in\TS{k}$ and $t\in[\sf T]$, can be obtained as a linear combination of the entries in $\tilde{ \*Y}_{\RS{j}}-\*Z_{\RS{j}}$. If it exists, we write this linear combination as
\begin{equation}\label{eq:linear}
X_{q}(t)= \mathcal{L}_{q,t} ( \tilde{ \*Y}_{\RS{j}}-\*Z_{\RS{j}}).
\end{equation}
Notice that the entries in $\tilde{ \*Y}_{\RS{j}}-\*Z_{\RS{j}}$ are themselves linear combinations of the inputs 
 inputs $\{X_{q}(t) \colon q\in\TS{k}, \; t\in[\sf T]\}$, and therefore the existence of linear functions $\mathcal{L}_{q,t}$ in \eqref{eq:linear} is equivalent to a given square matrix of channel coefficients being invertible. 
 
\item $(c)$ holds because $h (\tilde{\*Y}_{\mathcal{F}}|\mathbf{a}_r, \mathbf{a}_c, \*Y_{\mathcal{R}_{j}}, \mathcal{H}, E=0)$ is bounded and because  $\mathbb{P}(E=1)=1$. This latter fact holds because  $E=1$ whenever a specific square matrix  of the random channel coefficients is invertible (see $(b)$ above), which happens with probability 1 since the channel coefficients are independently  drawn from a continuous distribution; 
\item in $(d)$ we defined the tuple  $ {\*Z}_j' \triangleq (Z_{p}'(t)\colon p \in \RS{j},\; t \in [\sf T])$ and 
 \begin{equation}
 Z_p'(t) \triangleq Z_p(t) -  \sum_{q \in \TS{k}} H_{q,p}(t) \mathcal{L}_{p,t} ( \tilde{ \*Y}_{\RS{j}}-\*Z_{\RS{j}}),
\end{equation}
where $\mathcal{L}_{q,t}$ is from $(b)$;
\item in $(e)$ we used the independence of the noise from the channel coefficients and the fact that conditioning can only decrease differential entropy.

\end{itemize}

Finally, combining Eqs. \eqref{eq:p1}, \eqref{eqn:converse_part3}, and \eqref{eqn:converse_part4_2}, we obtain
\begin{IEEEeqnarray}{rCl}
 \lefteqn{  \sum_{p \in \RS{j}} \;\sum_{\ell \in [\tilde{\sf K}]\backslash\{j\}}  \sf R_{p, \ell} + \sum_{p \in [\sf K] \backslash (\TS{k} \cup \RS{j})}  \sf R_{p,k}} \quad \nonumber \\
 &\leq&  \sf r   \log \left(1+ \sf P \sf H_{\max}^2 (\sf K-\sf r )\right) + \epsilon_{\sf T} \nonumber \\
 & &+ \frac{1}{\sf T} h({\*Z}_{\mathcal{F}}')- \frac{1}{\sf T}h(\*Z_{\mathcal{F}}) + \tilde{\epsilon}_{\sf T}.\hspace{3cm}
\end{IEEEeqnarray}
Letting $\sf T \to \infty$ and $\sf P \to \infty$ establishes the desired inequality in the lemma, because $h({\*Z}_{\mathcal{F}}')/\sf T$ and $h(\*Z_{\mathcal{F}})/\sf T$ are both finite constants that do not depend on $\sf T$ nor $\sf P$, and both sequences $\epsilon_{\sf T}$ and $\tilde{\epsilon}_{\sf T}$ tend to 0 as $\sf T \to \infty$.
\end{IEEEproof}}

\lo{
\section{Proof that submatrices of $\hat{\mat{D}}_p$, for $p\in\mathcal{T}_2$, are  full-rank with probability 1}\label{app:Pr}

Consider any square sub-matrix $\hat{\mat{D}}_p'$ of $\hat{\mat{D}}_p$ and define the  function 
\begin{IEEEeqnarray}{rCl}
F\left(  \left\{\mat{S}_{p \textnormal{ mod } \sf r}^{(2,k)} \colon k \in [\tilde{\sf K}]\backslash \{2\}\right\} , \; \mathcal{G}_2 \right) \triangleq \det\left( \hat{\mat{D}}_p' \right),
\end{IEEEeqnarray}
which is a polynomial in the entries of the matrices $\left\{\mat{S}_{p \textnormal{ mod } \sf r}^{(2,k)} \colon k \in [\tilde{\sf K}]\backslash \{2\}\right\}$ and $\mathcal{G}_2$.
According to \eqref{eq:bigH}, \eqref{eq:def_matrix_G}, \eqref{eq:Gexp}, and \eqref{eq:set_G},
$F$ is a rational function in the entries of the matrices $\{\mat{H}_{p,q}\}$ and $\{\mat{S}_{\ell}^{(i, k)}\}$, where the polynomial in the denominator (which consists of  products of determinants of matrices  $\tilde{\mat{H}}^{(1,2)}$ and $\{\tilde{\mat{H}}^{(2,k)}\colon \; k\in[\tilde{\sf K}]\backslash \{2\}\}$) is bounded and non-zero by our assumption that all channel matrices $\tilde{\mat{H}}^{(i,k)}$ are invertible. The zero-set  of the rational function $F$ is thus of Lebesgue measure 0 unless $F$ is equal to the all-zero function. (This can be seen by noting that the zeros of $F$ are the zeros of the polynomial in its numerator, which have Lebesgue measure 0  except when the polynomial is the all-zero polynomial, i.e., when $F$ is the all-zero function.) 
  Since real and imaginary parts of all entries of matrices $\{\mat{H}_{p,q}\}$ and $\{\mat{S}_{\ell}^{(i,\bar k)}\}$ are drawn independently from continuous distributions, we conclude that  the function $F$ evaluates to 0 
with probability 0 (over the matrices  $\{\mat{H}_{p,q}\}$ and $\{\mat{S}_{\ell}^{(i, k)}\}$), except for the case where it is the all-zero function.

In the rest of this section, we show that $F$ is not the all-zero function, or equivalently that the determinant of $\hat{\mat{D}}_p'$ is non-zero for at least one realization of the random matrices. In fact, we  show the stronger statement that for the realizations 
\begin{subequations} \label{eq:simp}
\begin{IEEEeqnarray}{rCl}
\tilde{\mat{H}}^{(1,2)} & = &   \mat{Id}_{\sf T \sf r} \\
\tilde{\mat{H}}^{(2,k)}&= &\mat{Id}_{\sf T \sf r} , \quad k \in[\tilde{\sf K}]\backslash \{2\},
\end{IEEEeqnarray}
\end{subequations}
the determinant of $\hat{\mat{D}}_p'$  is non-zero with probability 1. To this end, notice that 
for the realizations in  \eqref{eq:simp}, for any distinct triple $(\bar i,\bar j,\bar k) \in [\tilde{\sf K}]^3$ with either $(\bar i,\bar k)=(1,2)$ or $\bar i=2$:
\begin{equation}
\tilde{\mat{G}}_{\bar j}^{(\bar i,\bar k)}=\tilde{\mat{H}}^{(\bar j,\bar k)} \mat{S}^{(\bar i,\bar k)},
\end{equation}
which implies that  for any $\bar{p}=(\bar i-1) \sf r + \bar{\ell}$ in group $\TS{\bar i}$ and $p'=(\bar j-1) \sf r +\ell$ in group $\TS{\bar j}$, for $\bar{i}$ and $\bar{j}$ as above:
\begin{equation}
\mat{G}_{p'}^{(\bar p, \bar k)}={\mat{H}}_{p', ( \bar k-1) \sf r +\bar{\ell}} \;  \mat{S}^{(\bar i, \bar k)}_{ \ell},
\end{equation}
 because $\mat{S}^{(\bar i,\bar k)}$ is diagonal and $\tilde{\mat{H}}^{(\bar j,\bar k)}$ consists of $\sf r^2$ blocks of $\sf T$-dimensional block matrices. As a consequence, for the realizations in \eqref{eq:simp}, the matrix $\hat{\mat{D}}_p$ is given by \eqref{eq:Matform2} on top of the next page. 
 \begin{figure*}[h!]
\lo{\begin{IEEEeqnarray}{rCl}
\lefteqn{ \hat{\mat{D}}_p\Big|_{\eqref{eq:simp}}} \nonumber\\
  & = &  \Bigg[\mat{S}^{(2,k)}_{p \textnormal{ mod } \sf r} \cdot  \prod_{\substack{ (\bar k, \bar p, p') \colon \bar p \in \TS{2} \\ \bar  k \in[\tilde{\sf K}] \backslash \{2\}, \\ p' \in [\sf{K}]\backslash  (\TS{2} \cup \TS{\bar k}) }} \left( \mat{G}_{p'}^{(\bar p,\bar k)} \right)^{{\alpha_{2,(\bar k,\bar p,p')}}} \cdot\hspace{-.6cm}  \prod_{\substack{ \bar p   \in \TS{1}, \\ p' \in[\sf K]\backslash \{ \TS{1} \cup \TS{2}\}}  }  \hspace{-0.6cm} 
\left(\mat{G}_{\bar p}^{(p',2)}\right)^{{\alpha_{2,(\bar p,p')}}}    \cdot \boldsymbol{1} \colon \ k \in [\tilde{\sf K}]\backslash\{2\} , \;\{ \alpha_{2,\bar p,p'}\}, \{\alpha_{2,(\bar k,\bar p,p')} \}\in [\eta]\Bigg] \nonumber\\
\label{eq:Matform1}\\
& = &  \Bigg[\mat{S}^{(2,k)}_{p \textnormal{ mod } \sf r} \cdot  \prod_{\substack{ (\bar k, \bar p, p') \colon \bar p \in \TS{2} \\  \bar k \in[\tilde{\sf K}] \backslash \{2\}, \\ p' \in [\sf{K}]\backslash  (\TS{2} \cup \TS{k}) }} \left(    \mat{H}_{p',(\bar k -1)\sf r + (\bar{p} \mod \sf r)} \cdot \mat{S}_{ p' \mod \sf r}^{(2,\bar k)}  \right)^{{\alpha_{2,(\bar k,\bar p,p')}}}
 \nonumber \\
 & &\qquad \qquad
 \cdot \hspace{-0.6cm} \prod_{\substack{ \bar p   \in \TS{1}, \\ p' \in[\sf K]\backslash \{ \TS{1} \cup \TS{2}\}}  } 
 \hspace{-0.6cm} \left(     \mat{H}_{\bar p, r + ({p}' \mod \sf r)}  \cdot \mat{S}_{p' \mod \sf r}^{\lfloor  p'/\sf r \rfloor,2 }\right)^{{\alpha_{2,(\bar p,p')}}}   \cdot \boldsymbol{1}  \colon \  k  \in [\tilde{\sf K}]\backslash\{2\} , \; \{\alpha_{2,\bar p,p'}\},\{\alpha_{2,(\bar k,\bar p,p')} \}\in [\eta]\Bigg]. \label{eq:Matform2}
\end{IEEEeqnarray}
\hrule}
\end{figure*}
In the following, we explain in detail that the  matrix in \eqref{eq:Matform2} has the same form as matrix $\mat{A}$ in Lemma~\ref{lma:lemma1} at the end of this section. Trivially, then also any square submatrix of  $\hat{\mat{D}}_p$ has the same form, which by Lemma~\ref{lma:lemma1} proves that  for the realizations in \eqref{eq:simp} the determinant of $\hat{\mat{D}}_p'$ is non-zero with probability $1$. 



To see that $\hat{\mat{D}}_p$ is of the form in \eqref{eq:lemma1_A}, notice that all matrices involved in \eqref{eq:Matform2} are diagonal, and their multiplications with an all-one vector from the right leads to a column-vector consisting of the non-zero entries of these diagonal matrices. More precisely, the random variables in row $t$ are given by the slot-$t$ channel coefficients $\{H_{q,p}(t)\}$ and the $t$-th diagonal elements of $\mat{S}_{\ell}^{(i,k)}$, which by definition are independent of each other and of all random variables in the other rows. Therefore, the matrix  \eqref{eq:Matform2} satisfies Condition~i) in Lemma~\ref{lma:lemma1}. To see that it also satisfies Condition~ii), notice that there is a one-to-one mapping between the columns of  $\hat{\mat{D}}_p$ and the parameter tuples $\vect{v} = (k, \{\alpha_{2,(\bar k, \bar p, p')}\}, \{\alpha_{2,(\bar p, p')}\})$ and that for any two distinct tuples  $\vect{v}^{(1)} = (k^{(1)}, \{\alpha^{(1)}_{2,(\bar k, \bar p, p')}\}, \{\alpha^{(1)}_{2,(\bar p, p')}\})$ and  $\vect{v}^{(2)} = (k^{(2)}, \{\alpha^{(2)}_{2,(\bar k, \bar p, p')}\}, \{\alpha^{(2)}_{2,(\bar p, p')}\})$ the exponents in the corresponding  columns differ because:
\begin{enumerate}
	\item If $\alpha^{(1)}_{2,(\bar p, p')} \neq \alpha^{(2)}_{2,(\bar p, p')}$, then $H_{\bar p, \sf r + (p' \textnormal{ mod } \sf r)}$ has different exponents in the two columns.
	\item If $\alpha^{(1)}_{2,(\bar k, \bar p, p')} \neq \alpha^{(2)}_{2,(\bar k, \bar p, p')}$, then $H_{p', (\bar k -1)\sf r + (\bar p \textnormal{ mod } \sf r)}$ has different exponents in the two columns.
	\item If $\alpha^{(1)}_{2,(\bar p, p')} = \alpha^{(2)}_{2,(\bar p, p')}$ and  $\alpha^{(1)}_{2,(\bar k, \bar p, p')} = \alpha^{(2)}_{2,(\bar k, \bar p, p')}$, but $k^{(1)} \neq k^{(2)}$, then both $\mat{S}^{(2,k^{(1)})}_{p \textnormal{ mod } \sf r}$ and $\mat{S}^{(2,k^{(2)})}_{p \textnormal{ mod } \sf r}$ have different exponents in the two columns.
        \end{enumerate}
        This concludes the proof. 
 \begin{lemma}[Lemma~1 in \cite{cadambe_interference_2009}] \label{lma:lemma1}
 Consider an $\sf M$-by-$\sf M$ square matrix $\textbf{A}$ with $i$-th row and $j$-th column entry \begin{equation} \label{eq:lemma1_A}
     a_{ij} = \prod_{\ell=1}^{\sf L} \left( X_i^{[\ell]}\right)^{\alpha_{ij}^{[\ell]}}, \qquad  i,j \in \sf M,
 \end{equation}
 for  random variables  $\{X_i^{[\ell]}\}_{\ell \in [\sf L]}$ and exponents 
 \begin{equation}
     {\boldsymbol\alpha}_{ij} \triangleq \left(\alpha_{ij}^{[1]}, \alpha_{ij}^{[2]}, \ldots, \alpha_{ij}^{[\sf L ]}\right) \in \mathbb{Z}^{+\sf L}.
 \end{equation} 
 If 
 \begin{enumerate}
     \item for any two pairs $(i,\ell) \neq (i', \ell')$ the conditional cumulative probability distribution $P_{ X_i^{[\ell]}|X_{i'}^{[\ell']}}$ is continuous; and
     \item any pair of vectors ${\boldsymbol\alpha}_{i,j} \neq {\boldsymbol\alpha}_{i,j'}$ for $i, j,j' \in [\sf M]$ with $j \neq j'$;
 \end{enumerate}
 then the matrix $\textbf{A}$ is full rank with probability 1.
 \end{lemma}

}

\bibliographystyle{IEEEtran}
\bibliography{IEEEabrv,references}

\end{document}